\documentclass{vgtc}                          

\ifpdf
  \pdfoutput=1\relax                   
  \usepackage{graphicx}                
  \DeclareGraphicsExtensions{.pdf,.png,.jpg,.jpeg} 
\else
  \usepackage{graphicx}                
  \DeclareGraphicsExtensions{.eps}     
\fi%

\graphicspath{{figures/}{pictures/}{images/}{./}} 

\usepackage{microtype}                 
\PassOptionsToPackage{warn}{textcomp}  
\usepackage{textcomp}                  
\usepackage{mathptmx}                  
\usepackage{times}                     
\usepackage{cite}                      
\usepackage{tabu}                      
\usepackage{booktabs}                  
\usepackage{balance}

\onlineid{0}

\vgtccategory{Research}

\vgtcinsertpkg

\title{VocabulARy replicated: comparing teenagers to young adults}

\author{
Maheshya Weerasinghe \thanks{e-mail: amw31@st-andrews.ac.uk} \\%
        \scriptsize University of St Andrews and\\ 
        \scriptsize University of Primorska  
\and Verena Biener \thanks{e-mail: verena.biener@hs-coburg.de}\\ %
Jens Grubert \thanks{e-mail: jens.grubert@hs-coburg.de}\\ %
        \scriptsize Coburg University of  \\
        \scriptsize Applied Sciences %
\and Jordan Aiko Deja \thanks{e-mail: jordan.deja@famnit.upr.si}\\ %
Nuwan T. Attygalle \thanks{e-mail: nuwan.attygalle@famnit.upr.si}\\ %
Karolina Trajkovska \thanks{e-mail: 89191037@student.upr.si}\\ %
\scriptsize University of Primorska 
\and Matjaž Kljun  \thanks{e-mail: matjaz.kljun@upr.si}\\ %
Klen Čopič Pucihar \thanks{e-mail: klen.copic@famnit.upr.si}\\ %
          \scriptsize University of Primorska and \\ \scriptsize Faculty of Information Studies Novo mesto\\ 
}

\abstract{
A critical component of 
user studies is gaining access to a representative sample of the population researches intend to investigate. Nevertheless, the vast majority of human-computer interaction (HCI) studies, including augmented reality (AR) studies, rely on convenience sampling. The outcomes of these studies are often based on results obtained from university students 
aged between 19 and 26 years. In order to investigate how the results from one of our studies are affected by convenience sampling, 
we replicated the AR-supported language learning study called VocabulARy with 24 teenagers, aged between 14 and 19 years. The results verified most of the outcomes from the original study. In addition, it also revealed that teenagers found learning significantly less mentally demanding compared to young adults, and completed the study in a significantly shorter time. All this at no cost to learning outcomes. 

} 

\CCScatlist{
  \CCScatTwelve{Human-centered computing}{Visu\-al\-iza\-tion}{Visu\-al\-iza\-tion techniques}{Treemaps};
  \CCScatTwelve{Human-centered computing}{Visu\-al\-iza\-tion}{Visualization design and evaluation methods}{}
}

\begin{document}

\maketitle


\section{Introduction}

Reproducibility of user studies is a well known problem affecting different scientific fields. 
For example, 
an attempt to replicate 100 studies in psychology, published within a year in three high-ranking psychology journals, revealed that only 36\% of results were replicated successfully~\cite{ReplicationPsi}. Furthermore, a survey including scientists from various disciplines (chemistry, biology, physics/engineering, medicine, earth and environment sciences, etc.) found that the vast majority of them (64\%-87\%) reported having problems replicating results from other studies~\cite{Hen2017}. Another problem is the lack of replication studies. A paper in the field of human-computer interaction (HCI) 
reports that only 3\% out of 891 studies attempted to replicate an earlier result~\cite{hornbaek2014once}. 
Within augmented, mixed and virtual reality (AR, MR, VR) research such studies are even more scarce~\cite{TutorialReplicationCrisis} with only a few examples available, such as~\cite{ARReplication, mohr2017adaptive}.

To conduct a robust user study, a careful consideration must be given to a
representative sample of the population researchers intend to investigate. 
However, the vast majority of HCI studies rely on convenience sampling. As a result, 
the conclusions made in the literature are often based on the university students, 
aged 19 through 30. This is also one of the recognised limitations of one of our previous user studies called VocabulARy\cite{weerasinghe2022vocabulary}. To investigate the effect of the convenience sampling on the results of the aforementioned study, we decided to run a replication study that targeted a different age group, aged 14 through 19.

VocabulARy is an AR system for learning words in a foreign language. For our studies we selected Japanese as a language to be learnt because it is uncommonly spoken by speakers of Indo-European languages. The prototype displays visual and audio AR annotations for objects in users' surroundings. For each object users can see two words -- an English word (first language) and Japanese translation (second language) -- and the audio pronunciation of the latter. 
In addition, the prototype displays a keyword and its visualisation to enhance memory retention. In our studies we compare the AR system to the non-AR tablet computer and on each we compare keywords to keywords together with its visual representation.  


\section{User study}
The study was conducted following the research method described in detail in the original paper~\cite{weerasinghe2022vocabulary}. The summary of the procedure and differences are explained hereafter. 

\subsection{Participants and procedure}
The study was completed by 24 participants aged 14 through 19 ($\overline{x} = 15.8$, $SD = 1.5$). Half of them (12) participated in the \textsc{ar} (5 female) and half in the \textsc{non-ar} condition (6 female). 
As we opted for a mixed design study, the between subject factor (i.e. \textsc{ar} and \textsc{non-ar}) could be studied at two different locations. The \textsc{non-ar} was studied at a scout camp, 
and the \textsc{ar} at the university as a part of a 
summer school. 
Since the participants came from different schools and different parts of the country, they represent a more varied sample compared to recruiting participants from one school or area only. 
All participants 
voluntarily took part in the study and the consent forms where acquired from their parents or legal guardians if they were younger than 18. 

At each location, 
we randomly selected the instruction mode to be used first (\textsc{keyword} vs \textsc{keyword+visualisation}).
Finally, the learning scenario was randomly selected (kitchen or office environment). All randomisations were counterbalanced. After the training session the participants were asked to remember 10 Japanese words in the learning scenario given. 

\subsection{Differences between the replication and original study}
Besides the age group (young adults \textsc{19-30} vs teenagers \textsc{14-19}) and the sampling method (convenience sampling among computer science students from one university vs sampling from a more varied group of high-school students), there were two other differences. 
In the replication study we did not capture the delayed recall data, and the maximal available time for \textsc{non-ar} condition was reduced from 15 to 5 minutes due to organisational limitations 
set by camp organisers. 

\begin{figure*}[ht]
	\centering 
	\includegraphics[width=2.1\columnwidth]{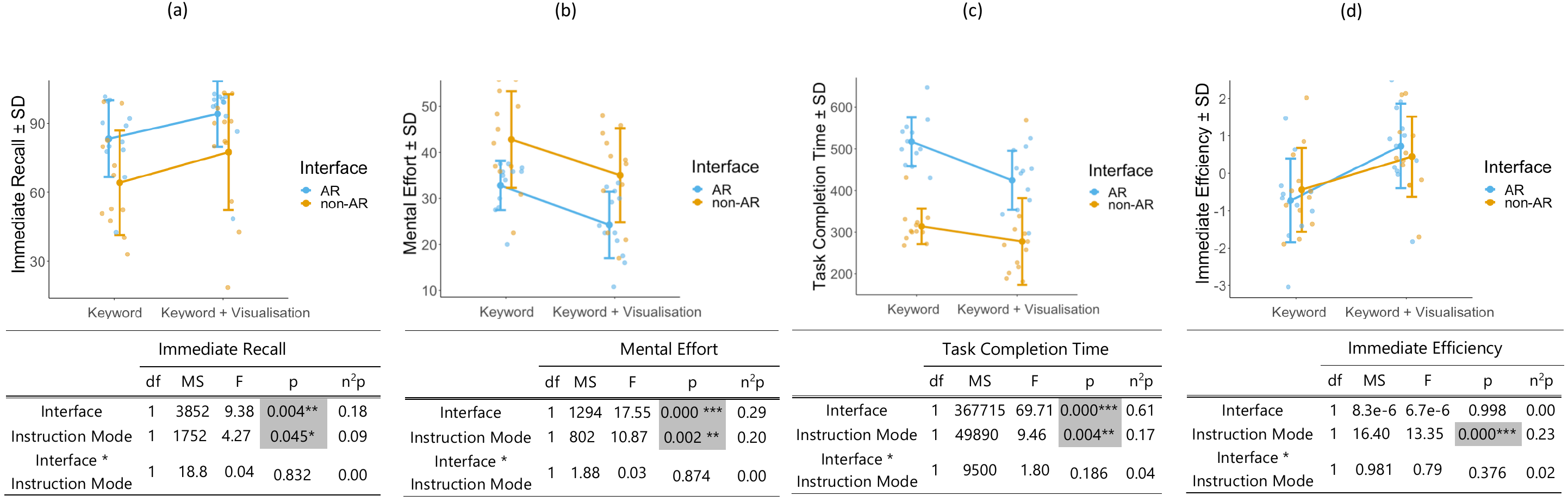}
 	\caption{Means with standard deviation and ANOVA results for: (a) immediate recall performance in percentage of correctly remembered words; (b) mental effort invested during the study; (c) task-completion-time in seconds; (d) learning efficiency.}
	\label{fig:set3}
\end{figure*}
\begin{figure*}[ht]
	\centering 
	\includegraphics[width=2.1\columnwidth]{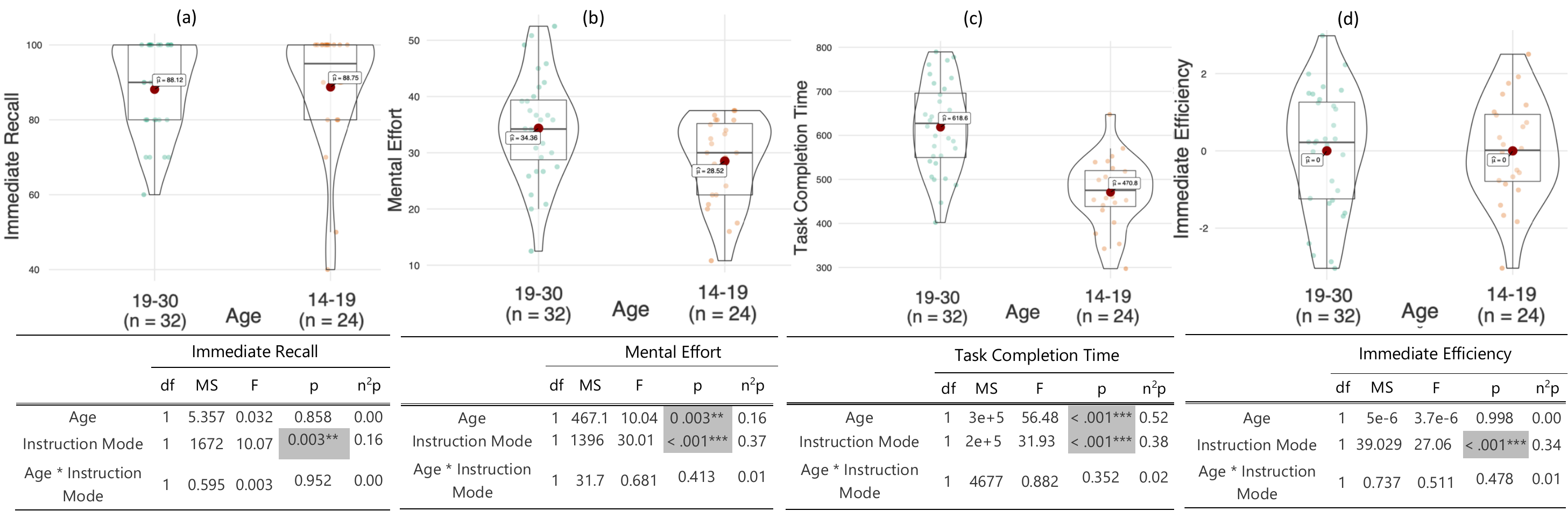}
 	\caption{Means with standard deviation and ANOVA results for \textsc{ar} condition analysing the effect of age. This analysis includes data from the original and replication study for: (a) immediate recall performance in percentage of correctly remembered words; (b) mental effort invested during the study; (c) task-completion-time in seconds; (d) learning efficiency.}
	\label{fig:set4}
\end{figure*}

\section{Results}
All participants managed to successfully complete the study. The dependent variables (Immediate Recall (how many words could be successfully recalled after the study), Mental Effort (measured with the NASA-TLX questionnaire), Task Completion Time (time needed to learn all the words), Learning Efficiency (the ratio of performance to the difficulty of the learning task)) as well as all statistical analysis carried out are described in detail in the original paper~\cite{weerasinghe2022vocabulary}. 

We conducted a mixed design analysis with 2 \textsc{age} (\textsc{14-19} vs \textsc{19-30}) x 2 \textsc{interface} (\textsc{non-ar} vs \textsc{ar}) x 2 \textsc{instruction mode} (\textsc{keyword} vs \textsc{keyword+visualisation}) conditions. 
\textsc{age} and \textsc{interface} conditions were analysed as between-subject factors, all others were analysed as within-subject factors. 
The results are organised according to dependent variables and divided into \textit{Replication study} and \textit{Replication and original study} subsections.




\subsection{Immediate Recall}

\textbf{Replication study:} The mean values of immediate recall and the ANOVA results for the \textsc{interface} (\textsc{ar} and \textsc{non-ar}) and the \textsc{instruction mode} (\textsc{keyword} and \textsc{keyword+visualisation}) conditions are shown in Figure~\ref{fig:set3}a. 

A significant main effect of the \textsc{interface} on immediate recall could be detected ($F(1,44) = 9.38$, $p < 0.05$, $n^{2}p = 0.18$). Immediate recall scores were significantly better in \textsc{ar} condition ($\overline{x} = 88.8\%$, $SD = 24.48$) compared to the \textsc{non-ar} condition ($\overline{x} = 70.8\%$, $SD = 16.24$).

Also, a significant main effect of \textsc{instruction mode} on immediate recall performance could be detected ($F(1,44) = 4.27$, $p < 0.05$, $n^{2}p = 0.09$). Immediate recall scores in \textsc{keyword+visualisation} condition ($\overline{x} = 85.8\%$, $SD = 21.85$) were significantly better than in \textsc{keyword} condition ($\overline{x} = 73.8\%$, $SD = 21.83$). No significant effect could be found between the \textsc{interface} and \textsc{instruction mode} ($F(1,44) = 0.05$, $p > 0.05$, $n^{2}p < 0.001$).


\textbf{Replication and original study for AR:} The distribution of the data and the ANOVA results for immediate recall focusing on the effect of \textsc{age} 
are shown in Figure~\ref{fig:set4}a.
Statistical analysis 
showed no significant effect of the \textsc{age} on participants' immediate recall ($F(1,52) = 0.032$, $p > 0.05$, $n^{2}p < 0.001$). Also, no significant interaction effect could be found between the \textsc{age} and the \textsc{instruction mode} conditions ($F(1,52) = 0.003$, $p > 0.05$, $n^{2}p < 0.001$). 

\subsection{Mental Effort}
\label{subsec:ME}
\textbf{Replication study:} The mean values of the mental effort experienced during the task 
and the ANOVA results 
are shown in Figure~\ref{fig:set3}b.

A significant effect of the \textsc{interface} on mental effort could be detected ($F(1,1) = 17.54$, $p < 0.001$, $n^{2}p = 0.26$). It was significantly lower for the \textsc{ar} ($\overline{x} = 28.5$, $SD = 10.87$) compared to the \textsc{non-ar} condition ($\overline{x} = 38.9$, $SD = 7.61$). Also, a significant effect of the \textsc{instruction mode} on mental effort could be detected ($F(1,1) = 10.87$, $p < 0.01$, $n^{2}p = 0.29$). In \textsc{keyword+visualisation} ($\overline{x} = 29.6$, $SD = 10.25$) it was significantly lower than in \textsc{keyword} condition ($\overline{x} = 37.8$, $SD = 9.61$). No significant effects were found between the \textsc{interface} and \textsc{instruction mode} ($F(1,1) = 0.03$, $p > 0.05$, $n^{2}p < 0.001$).


\textbf{Replication and original study for AR:} The distribution of the data and the ANOVA results for mental effort focusing on the effect of \textsc{age}
are shown in Figure~\ref{fig:set4}b. A significant effect of the \textsc{age} on mental effort could be detected ($F(1,52) = 10.04$, $p < 0.05$, $n^{2}p = 0.16$). Mental effort was significantly lower in the \textsc{14-19} age group ($\overline{x} = 28.5$, $SD = 7.61$) compared to the \textsc{19-30} age group ($\overline{x} = 34.4$, $SD = 9.17$). However, no significant interaction effect could be found between the \textsc{age} and the \textsc{instruction mode} ($F(1,52) = 0.681$, $p > 0.05$, $n^{2}p = 0.01$). 

\subsection{Task Completion Time}
\textbf{Replication study:} The mean values of task completion time for all study conditions 
are shown in Figure~\ref{fig:set3}c. The data 
is analysed using a between-within subjects ANOVA on the 20\% trimmed means~\cite{mair2020robust}. 

A significant effect of the \textsc{interface} on task completion time could be detected ($F(1,1) = 69.71$, $p < 0.001$, $n^{2}p = 0.61$). The completion time was significantly lower for the \textsc{non-ar} condition ($\overline{x} = 296 s$, $SD = 80 s$) compared to the \textsc{ar} condition ($\overline{x} = 475 s$, $SD = 79 s$).

Also, a significant effect of \textsc{instruction mode} on task completion time could be detected ($F(1,1) = 9.46$, $p < 0.01$, $n^{2}p = 0.17$). The \textsc{keyword+visualisation} ($\overline{x} = 351 s$, $SD = 110 s$) resulted in a significantly lower completion time than \textsc{keyword} ($\overline{x} = 416 s$, $SD = 115 s$). There was no significant effect between the \textsc{interface} and \textsc{instruction mode} ($F(1,1) = 1.80$, $p > 0.05$, $n^{2}p = 0.04$).


\textbf{Replication and original study for AR:} The distribution of the data and the ANOVA results for task completion time focusing on the effect of \textsc{age}
are shown in Figure~\ref{fig:set4}c. A significant effect of \textsc{age} on task completion time could be detected ($F(1,52) = 56.48$, $p < 0.001$, $n^{2}p = 0.52$). The task completion time was significantly lower for \textsc{14-19} age group ($\overline{x} = 470.8$, $SD = 79.24$) than the \textsc{19-30} age group ($\overline{x} = 618.6$, $SD = 101.11$). No significant effect could be found between the \textsc{age} and \textsc{instruction mode} ($F(1,52) = 0.882$, $p > 0.05$, $n^{2}p = 0.02$) conditions. 


\subsection{Learning Efficiency}
\textbf{Replication study:} The average learning efficiency for immediate recall across study conditions are shown in Figure~\ref{fig:set3}d. The data 
is analysed using a between-within subjects ANOVA on the 20\% trimmed means~\cite{mair2020robust}.

Statistical analysis 
showed no significant effect of the \textsc{interface} on learning efficiency for immediate recall ($F(1,1) = 6.8e-6$, $p > 0.05$, $n^{2}p < 0.001$). A significant effect of the \textsc{instruction mode} condition on learning efficiency for immediate recall could be detected ($F(1,1) = 13.35$, $p < 0.001$, $n^{2}p = 0.23$). The learning efficiency was significantly higher in \textsc{keyword+visualisation} ($\overline{x} = 0.585$, $SD = 1.08$) compared to the \textsc{keyword} ($\overline{x} = -0.584$, $SD = 1.10$) condition. There was no significant effect between the \textsc{interface} and \textsc{instruction mode} for immediate recall ($F(1,1) = 0.79$, $p > 0.05$, $n^{2}p < 0.05$).


\textbf{Replication and original study for AR:}
The distribution of the data and the ANOVA results for learning efficiency focusing on the effect of \textsc{age}
are shown in Figure~\ref{fig:set4}d. Statistical analysis 
showed no significant effect of \textsc{age} for learning efficiency for immediate recall ($F(1,52) < 0.001$, $p > 0.05$, $n^{2}p < 0.001$). Also, no significant interaction effects could be found between the \textsc{age} and \textsc{instruction mode} ($F(1,52) = 0.511$, $p > 0.05$, $n^{2}p = 0.01$) conditions. 

\section{Discussion and Conclusion}
A comparison of replication and original study results shows that there is little difference between the two. The results for statistical tests for the dependent variables such as immediate recall, mental effort and immediate efficiency lead to the same conclusions. However, this was not the case for the time taken to complete the task. The results of both studies agree on the effect of \textsc{instruction mode}, whilst show the opposite in case of \textsc{interface} condition. For this condition the completion time is significantly lower for the \textsc{non-ar} compared to the \textsc{ar} condition. This is probably due to different time constraint for the \textsc{non-ar} and \textsc{ar} conditions. 

Furthermore, there is an observable difference in the significance levels that were detected for all the variables. As the sample size in the replication study was substantially smaller (i.e. $n = 24$ vs. $n = 32$ in the original study) one would expect higher or similar p-values. This was indeed observed for all p-values except for the p-value of Immediate Recall for \textsc{interface} condition (i.e. $p = 0.004$ vs. $ p = 0.01$ in original study). 
We hypothesise that the \textsc{non-AR} condition in the replication study was tested outside the laboratory where the researchers did not have a complete control over the environment. Thus, various disruptions could occur, such as noise, people walking into the room, the presence of other observers. In addition, it is important to note that the replication study did not capture data for delayed recall, thus this part was not presented here. 

Finally, the results analysing the effect of \textsc{age} condition showed that teenagers found the study significantly less mentally demanding and completed it in a significantly shorter time also in \textsc{ar} that had the same time constraints as the original study. However, despite overall better performance in immediate recall and learning efficiency, no significance was detected. Why this is the case remains to be answered.


\acknowledgments{
Authors would like to thank Nikola Kovačević, Nina Chiarelli and Ana Zalokar for their help with the study. This research was supported by European Commission through the InnoRenew CoE project (Grant Agreement 739574) under the Horizon2020 Widespread-Teaming program and the Republic of Slovenia (investment funding of the Republic of Slovenia and the European Union of the European Regional Development Fund). The work was also supported by the Slovenian research agency (program no. BI-DE/20-21-002, P1-0383, J1-9186, J1-1715, J5-1796, and J1-1692).
}

\balance 

\bibliographystyle{abbrv-doi}

\bibliography{template}
\end{document}